\documentclass[10pt]{article}
\usepackage{graphicx}
\usepackage{dcolumn}
\usepackage{bm}

\begin{document}

\title{Analytical Study of the Spin Projection Operator}
\author {Zhi-Qiang Shi\thanks{E-mail address: zqshi@snnu.edu.cn}\\{\it \small Department of Physics, Shaanxi Normal University, Xi'an 710062, P. R. China}}

\date{}
\maketitle

\begin{abstract}

We present an analytical study of the spin projection operator of the spin state and the helicity state. It is pointed out emphatically that the former is Lorentz covariant, while the latter is a nonrelativistic two-component operator. In the special case of $\bm p=p_z$, the helicity state and the spin state are formally identical. However, even so their spin projection operators are still different and so the helicity state is not a special case of the spin state. This makes the spinor is the degenerate state of the two different spin projection operators. The calculation on the lifetime of polarized muons shows that this difference will inevitably lead to the left-right polarization-dependent lifetime asymmetry.

PACS numbers: 03.65.Pm, 11.80.Cr, 12.15.Ji

Keywords: Relativistic wave equations, helicity and invariant amplitudes, applications of electroweak models to specific processes.

\end{abstract}

\section{Introduction}\label{1}\noindent

``The polarization asymmetry will play a central role in precise tests of the standard model.'' \cite{1} The SLD Collaboration \cite{2,3} and the SLAC E158 Collaboration \cite{4,5} have precisely measured the polarization asymmetry in neutral weak currents processes mediated by $Z^0$ exchange and got a definite proof. Their results show that the integrated cross section of left-handed (LH) polarized electrons is greater than that of right-handed (RH) ones in polarized electron-positron collisions and polarized electron-electron M{\o}ller scattering. It implies that the weak coupling strength of LH chirality states to $Z^0$ boson is greater than that of RH ones. This polarization asymmetry should be much more evident in charged weak currents processes mediated by $W^\pm$ exchange, because the weak coupling strength of RH chirality states to the $W^\pm$ boson is equal to zero. The fermion¡¯s decay processes are caused by charged weak currents and the lifetime is the reciprocal of the cross section (i.e., decay rate), and so it should also be the polarization asymmetry, which is called the lifetime asymmetry \cite{6,7,8,9}.

However, the study of the lifetime asymmetry has not aroused more attention. This implies that the difference between the spin and the helicity states, or the chirality and the helicity, still need to be more fully investigated theoretically. The spin and the helicity states are two kinds of the common-used spinor wave functions in Relativistic Quantum Mechanics. In the previous papers \cite{7,8}, the differences and the relations between them have been discussed. In this paper, we will focus on analytical study of their spin projection operators, which has hardly been studied before. It not only enables us to profound understand the essential difference between two kinds of spinor, but also can more intuitive, more rigorously prove the lifetime asymmetry. The paper is organized as follows. In Sec.2, the spin state and its spin projection operator are introduced concisely. In Sec.3, the relations between the spin projection operator of the helicity and the spin states are carefully discussed. In Sec.4 in order to investigate the polarization asymmetry in charged weak current, the lifetime asymmetry is proven by using the spin projection operator. Finally, we briefly summarize the study above in Sec.5.

\section{The spin projection operator of the spin states}\label{2}\noindent

The famous physicist A.~S.~Goldhaber and M.~Goldhaber pointed out emphatically that the chirality and the helicity are two completely different concepts \cite{10}. The chirality is the projection of a four-component Dirac spinor $u_s(p)$ on the chirality operator $\gamma_5$ and in Pauli metric, LH chirality state $u_{_{LS}}(p)$ and RH chirality state $u_{_{RS}}(p)$ are defined as, respectively,
\begin{equation}\label{eq:1}
  u_{_{LS}}(p)=\!\frac{1}{2}(1+\gamma_5)u_s(p),\quad  u_{_{RS}}(p)=\!\frac{1}{2}(1-\gamma_5)u_s(p).
\end{equation}
Their projection operators are, respectively,
\begin{equation}\label{eq:2}
    \rho_{_L}=\frac{1}{2}(1+\gamma_5),\quad \rho_{_R}=\frac{1}{2}(1-\gamma_5).
\end{equation}
The Standard Model of particle physics is a chiral gauge theory, in which all of fundamental fermions are divided into two classes, LH and RH chirality state, and they construct the weak interaction Lagrangian density with different gauge transformations. RH chirality states have zero weak isospin and are only present in neutral weak currents. In charged weak currents, all fermions are in the LH chirality states while all antifermions are in the RH ones, so that the parity violation reaches its maximum. Thus, LH and RH in the model all refer to the chirality states.

The plane wave solutions of Dirac equation have many kinds, but the most commonly used positive and negative energy solution, in momentum representation for a given 4-momenta $p$ and mass $m$, are respectively
\begin{equation}\label{eq:3}
  u_s(p)=\sqrt{\frac{E_p+m}{2m}}\left(\!\begin{array}{c}\varphi_s\\\frac{\displaystyle \bm \sigma\cdot\bm p}{\displaystyle E_p+m}\varphi_s\;\end{array}\!\right)\;,
\end{equation}
\begin{equation}\label{eq:4}
  v_s(p)=\sqrt{\frac{E_p+m}{2m}}\left(\!\begin{array}{c}\frac{\displaystyle \bm \sigma\cdot\bm p}{\displaystyle E_p+m}\varphi_s\\\varphi_s\end{array}\!\right)\;,
\end{equation}
where $E_p>0,\; s=1, 2$ and $\varphi_s$ are Pauli spin wave functions. The solutions are the eigenstates of the Pauli-Lubanski operator $\frac{\omega(p)\cdot e}{m}$ with eigenvalues $\pm 1$. The $\frac{\omega(p)}{m}$ is the Pauli-Lubanski covariant spin vector and $e$ is the four-polarization vector in the form
\begin{equation}\label{eq:6}
  e_{\alpha}=\left\{\begin{array}{ll}\bm e^0+\frac{\displaystyle\bm p\;(\bm p\cdot\bm e^0)}{\displaystyle m(E_p+m)}\;,\quad &(\alpha=1,2,3)\\
  i\;\frac{\displaystyle\bm p\cdot\bm e^0}{\displaystyle m}\;,\quad &(\alpha=4)\end{array}\right.
\end{equation}
where $\bm e^0$ is equal to the unit vector of $z$ axis and in the rest frame, $e^0=(\bm e^0,0)=(0,0,1,0)$.

Because operator $\frac{\omega(p)\cdot e}{m}$ is formed by the projection of the Pauli-Lubanski covariant spin vector $\frac{\omega(p)}{m}$ on the four-polarization vector $e$, the solutions  Eqs.~(\ref{eq:3}) and (\ref{eq:4}) directly connected with the chirality are known as the relativistic spin states, or the spin states for short. Of course, they are different from the spin states in quantum mechanics.

The spin projection operators of the spin states are
\begin{equation}\label{eq:7}
  \rho_s=\frac{1}{2}(1\pm i\;\gamma_5\;\gamma\cdot e).
\end{equation}
The plus sign refers to $s=1$ and the minus sign to $s=2$. When $E\gg m$ (in the ultrarelativistic limit), it can be written as
\begin{equation}\label{eq:8}
    \rho_s(E\gg m)=\frac{1}{2}(1\mp \gamma_5).
\end{equation}

\section{The spin projection operator of the helicity states}\label{3}\noindent

The helicity is the projection of fermion's spin angular momentum on the direction of its momentum. Therefore, the polarization of fermions in flight must be described by the helicity states which are closely related to directly observable quantities experimentally. The helicity states are another important plane wave solutions of Dirac equation \cite{11}. The helicity states read
\begin{equation}\label{eq:9}
  u_{_h}(p)=\sqrt{\frac{E_p+m}{2m}}\left(\!\begin{array}{c}\varphi_{_h}\\
  \frac{\displaystyle h|\bm p|}{\displaystyle E_p+m}\varphi_{_h}
  \end{array}\right),
\end{equation}
where $\varphi_{_h}$ is the eigenstate of the helicity operator,
\begin{equation}\label{eq:10}
  \frac{\bm \sigma\cdot\bm p}{|\bm p|}\;\varphi_{_h}=h\;\varphi_{_h},\quad h=\pm 1
\end{equation}
with
\begin{equation}\label{eq:11}
    \varphi_{_{h=+1}}=\left(\!\begin{array}{c}\cos\frac{\theta}{2}\;e^{-i\frac{\phi}{2}}\\
    \sin\frac{\theta}{2}\;e^{i\frac{\phi}{2}}
    \end{array}\right),\;
    \varphi_{_{h=-1}}=\left(\!\begin{array}{c}-\sin\frac{\theta}{2}\;e^{-i\frac{\phi}{2}}\\
    \cos\frac{\theta}{2}\;e^{i\frac{\phi}{2}}
    \end{array}\right),
\end{equation}
where $\theta$ and $\phi$ are the polar angles of momentum $\bm p$ in polar-coordinates. The state with $h=+1$ is the RH helicity state while the state with $h=-1$ is the LH one. The spin projection operators of the helicity states are \cite{12}
\begin{equation}\label{eq:14}
  \rho_{_h}=\frac{1}{2}\left(1+h\frac{\displaystyle \bm \Sigma\cdot \bm p}{\displaystyle |\bm p|}\;\right)\;.
\end{equation}

Taking the simplest case of $\bm p: \bm p=p_z$, which does not lose the universality of problem, we have the LH helicity state $u_{_{Lh}}$ and the RH helicity state $u_{_{Rh}}$, respectively
\begin{eqnarray}
  u_{_{Lh}}(p)&=&\sqrt{\frac{E_p+m}{2m}}\left(\!\begin{array}{c}\varphi_2\\
  \frac{\displaystyle {-|\bm p|\;\varphi_2}}{\displaystyle {E_p+m}}\end{array}\!\right),\label{eq:15}\\
  u_{_{Rh}}(p)&=&\sqrt{\frac{E_p+m}{2m}}\left(\!\begin{array}{c}\varphi_1\\
  \frac{\displaystyle {|\bm p|\;\varphi_1}}{\displaystyle {E_p+m}}\end{array}\!\right).\label{eq:16}
\end{eqnarray}
Considering $\bm e=\frac{E}{m}\bm e^0$, $e_4=i\frac{E}{m}\beta$, $\beta$ is the velocity of fermions, and $\bm \Sigma=-i\gamma_5\gamma_4\bm \gamma$, the $\rho_{_h}$  can be expressed as
\begin{equation}\label{eq:17}
  \rho_{_{h}}=\frac{1}{2}(1\mp i\;\frac{m}{E}\;\gamma_5\;\gamma_4\;(\gamma\cdot e)\mp \gamma_5\;\beta).
\end{equation}
The plus sign refers to the LH helicity state and the minus sign to the RH one. Comparing with the $\rho_s$, one can see that there exists a new additional term containing $\beta$ in the $\rho_{_h}$. In this simple case, obviously, the helicity state and the spin state are formally identical, i.e.
\begin{equation}\label{eq:18}
    u_1(p)=u_{_{Rh}}(p),\quad u_2(p)=u_{_{Lh}}(p).
\end{equation}
However, even so their spin projection operators are still different. Hence, the helicity state absolutely is not a special case of the spin state.

In brief, for massive free fermions, the spin projection operator of  the helicity, the chirality and the spin states is completely different. One can see from Eqs.~(\ref{eq:14}), (\ref{eq:2}) and (\ref{eq:7}) that the spin projection operator is a nonrelativistic two-component operator for the helicity state, Lorentz invariant for the chirality one and Lorentz covariant for the spin one \cite{13}. Especially, the spin state and its spin projection operator are both helicity degenerate because they are both independent of $h$.

When $\bm p=p_z$, one can see from Eqs.~(\ref{eq:18}), (\ref{eq:17}) and (\ref{eq:7}) that the helicity degeneracy of the spin state disappears, but that of its spin projection operator still exists. It makes the same spinor can have two different spin projection operators, i.e. the spin projection operators of the spinor are degenerate.

When $m=0$ or $E\gg m$, the helicity, the chirality and the spin  states are completely identical and it can be seen from Eqs.~(\ref{eq:8}) and (\ref{eq:17}) that the spin projection operators of the spin and the helicity states are reduced to that of the chirality states, like Eq.~(\ref{eq:2}), so that the degeneracy of the spin projection operators disappears.

\section{The lifetime of polarized muons in flight}\label{4}\noindent

The negative muon decay can be written as
\begin{equation}\label{eq:19}
    \mu^-\longrightarrow e^{-}+\overline\nu_e+\nu_\mu.
\end{equation}
The lowest order decay rate or lifetime $\tau$ for muon decays, based on the perturbation theory of weak interactions, is given by
\begin{equation}\label{eq:20}
  \tau^{-1}_s=\frac{1}{(2\pi)^5}\;\frac{G^2}{2}\;\frac{m_\mu m_e}{E_p}\!\int\!\frac{d^3 q}{E_q}\;d^3 k\;d^3 k'\;\delta^4(p-q-k-k')\;M^2_s,
\end{equation}
where $m_\mu$ is muon mass and $m_e$ is electron mass. If the muons are polarized and if we do not observe the polarization of final-state fermions, then the transition matrix element is given by summing over all final fermion spins:
\begin{equation}\label{eq:21}
  M^2_s=\sum_{s',r,r'=1}^2\left[\;\overline{u}_{s'}(q)\;\gamma_\lambda\;(1+\gamma_5)\;v_{r'}(k')\;\right]^2\;
        \left[\;\overline{u}_r(k)\;\gamma_\lambda\;(1+\gamma_5)\;u_s(p)\;\right]^2.
\end{equation}
where $p$, $q$, $k$ and $k'$ are 4-momenta, while $s$, $s'$, $r$ and $r'$ are spin indices for $\mu$, $e$, $\nu_\mu$ and $\bar{\nu}_e$, respectively.

As mentioned in most literatures and textbooks \cite{13,14,15,16,17}, based on the spin projection operators of the spin state, Eq.~(\ref{eq:7}), we have
\begin{equation}\label{eq:22}
  u_s(p)\overline{u}_s(p)=\rho_s\;\Lambda_+(p)=\frac{1}{2}(1\pm i\gamma_5\gamma\cdot e)\frac{-i\gamma\cdot p+m_\mu}{2\;m_\mu},
\end{equation}
where $\Lambda_+(p)$ is the energy projection operator. Substituting Eq.~(\ref{eq:22}) into (\ref{eq:21}) and applying the trace theorems we obtain
\begin{equation}\label{eq:23}
  M_s^2=\frac{8\;{\cal F}_s}{m_\mu m_e E_k E_{k'}},
\end{equation}
where the decay amplitude
\begin{equation}\label{eq:24}
  {\cal F}_s=(p\cdot k')\;(q\cdot k)\mp m_\mu(e\cdot k')\;(q\cdot k)={\cal F}\mp m_\mu\;{\cal F}_e.
\end{equation}
Substituting Eq.~(\ref{eq:23}) into Eq.~(\ref{eq:20}), one has
\begin{equation}\label{eq:25}
  \tau^{-1}_s=\frac{4\;G^2}{(2\pi)^5}\frac{1}{E_p}\int\!\frac{d^3 q}{E_q}\frac{d^3 k}{E_{k}}\frac{d^3k'}{E_{k'}}\delta^4(p-q-k-k'){\cal F}_s.
\end{equation}

The integration to the right of $E_p^{-1}$ in Eq.~(\ref{eq:25}) is a Lorentz scalar and we can calculate the muon lifetime in its rest frame. Neglecting electron mass and taking $\bm p=p_z$, one easily verifies
\begin{equation}\label{eq:26}
    \int\!\frac{d^3 q}{E_q}\frac{d^3k}{E_{k}}\frac{d^3k'}{E_{k'}}\delta^4(p-q-k-k'){\cal F}_e
    =\int\!\frac{d^3 q}{E_q}\frac{d^3k}{E_{k}}\frac{d^3k'}{E_{k'}}\delta^4(p^0-q-k-k'){\cal F}_e^0=0,
\end{equation}
and
\begin{equation}\label{eq:27}
    \tau^{-1}_0=\frac{4\;G^2}{(2\pi)^5}\frac{1}{m_\mu}\int\!\frac{d^3 q}{E_q}\frac{d^3 k}{E_{k}}\frac{d^3k'}{E_{k'}}\delta^4(p^0-q-k-k'){\cal F}^0
    =\frac{G^2m^5_\mu}{192\;\pi^3},
\end{equation}
where ${\cal F}_e^0\;({\cal F}^0)$ is the ${\cal F}_e\;({\cal F})$ in the muon rest frame
\begin{equation}\label{eq:28}
    {\cal F}_e^0=(e^0\cdot k')\;(q\cdot k),\quad e^0=(0,0,1,0)
\end{equation}
\begin{equation}\label{eq:29}
  {\cal F}^0=(p^0\cdot k')(q\cdot k).\quad p^0=(0,0,0,im_\mu)
\end{equation}
The $E_p^{-1}$ in Eq.~(\ref{eq:25}) reflects the time dilation effect and thus in an arbitrary frame the muon lifetime is given by
\begin{equation}\label{eq:30}
    \tau_s=\tau=\frac{\tau_{_0}}{\sqrt{1-\beta^2}}.
\end{equation}

On the other hand, owing to the degeneracy of the spin projection operators, the spin projection operator of helicity state, Eq.~(\ref{eq:17}), can also be chosen. For LH helicity, instead of Eq.~(\ref{eq:22}), we have
\begin{equation}\label{eq:31}
  u_2(p)\overline{u}_2(p)=u_{_{Lh}}(p)\overline{u}_{_{Lh}}(p)=\rho_{_{h=-1}}\;\Lambda_+(p)
  =\frac{1}{2}(1+i\frac{m_\mu}{E_p}\;\gamma_5\;\gamma_4\;(\gamma\cdot e)+\gamma_5\;\beta)\frac{-i\gamma\cdot p+m_\mu}{2\;m_\mu}
\end{equation}
Since $(\gamma\cdot e)(\gamma\cdot p)=-(\gamma\cdot p)(\gamma\cdot e)=-(\gamma\cdot p^0)(\gamma\cdot e^0)=-im_\mu\;\gamma_4(\gamma\cdot e^0)$, we get
\begin{equation}\label{eq:32}
     u_{_{Lh}}(p)\overline{u}_{_{Lh}}(p)=\frac{-i}{4m_\mu}[(1+\gamma_5\beta)(\gamma\cdot p)+\frac{m^2_\mu}{E_p}\;\gamma_5\;(\gamma\cdot e^0)]
                                         +\frac{1}{4}[1+i\frac{m_\mu}{E_p}\;\gamma_5\;\gamma_4\;(\gamma\cdot e)+\gamma_5\;\beta].
\end{equation}
Owing to the trace of a product of an odd number of $\gamma$-matrices vanishes, the second term of the above equation will be deleted in trace operational process. Therefore we obtain
\begin{equation}\label{33}
    {\cal F}_{_{Lh}}=(1+\beta)(p\cdot k')(q\cdot k)+\frac{m^2_\mu}{E_p}(e^0\cdot k')(q\cdot k)=(1+\beta){\cal F}+ \frac{m^2_\mu}{E_p}\;{\cal F}^0_e.
\end{equation}
Comparing with Eq.~(\ref{eq:24}) and considering Eq.~(\ref{eq:26}) we obtain the lifetime of LH polarized muons
\begin{equation}\label{eq:34}
    \tau_{_{Lh}}=\frac{\tau}{1+\beta}.
\end{equation}

Similarly, for RH polarized muons we have
\begin{equation}\label{eq:35}
    u_1(p)\overline{u}_1(p)=u_{_{Rh}}(p)\overline{u}_{_{Rh}}(p)=\rho_{_{h=+1}}\;\Lambda_+(p)
    =\frac{1}{2}(1-i\frac{m_\mu}{E_p}\;\gamma_5\;\gamma_4\;(\gamma\cdot e)-\gamma_5\;\beta)\frac{-i\gamma\cdot p+m_\mu}{2m_\mu}.
\end{equation}
Then the lifetime is given by
\begin{equation}\label{eq:36}
    \tau_{_{Rh}}=\frac{\tau}{1-\beta}.
\end{equation}

It shows that for polarized muons in flight, if we choose the spin projection operator of the spin state to performing calculation, it will result in Eq.~(\ref{eq:30}) which does not reveal any polarization asymmetry; if we choose the spin projection operator of the helicity state, it will result in Eqs.~(\ref{eq:34}) and (\ref{eq:36}) in which the lifetime is the left-right polarization asymmetry. The lifetime asymmetry is expressed by
\begin{equation}\label{eq:37}
    A=\frac{\tau_{_{Rh}}-\tau_{_{Lh}}}{\tau_{_{Rh}}+\tau_{_{Lh}}}=\beta.
\end{equation}

Furthermore, this conclusion is also valid for all fundamental fermions in the decays under weak interactions. It means that the lifetime of RH polarized fermions is always greater than that of LH ones in any one of inertial systems in which fermions are in flight with a same speed. Detailed analysis showed that all of the existing experiments of measuring lifetime can not be used to directly prove the lifetime asymmetry, but the SLD and the E158 experiments have indirectly proven the lifetime asymmetry \cite{18}.

\section{Conclusion}\label{5}\noindent

Because the chirality is the projection of the spin state on the chirality operator $\gamma_5$, the difference between the chirality and the helicity can be attributed to the difference between the spin states and the helicity states. Their most fundamental difference is they have different spin projection operator. When $\bm p=p_z$, their spin projection operators are degenerate, i.e. a spinor has two different spin projection operators, one is the spin projection operator of the spin state and another is that of the helicity one. We should choose which one? The latter should be chosen because the helicity is an experimentally observable quantity.

It is easy to see that the relation between the spin state and the helicity state is similar to the relation between neutrino flavour eigenstate and mass eigenstate. The flavour eigenstates enter into the expression of the weak interaction Lagrangian, so they are also called the weak eigenstates, while the mass eigenstates describe the propagation of neutrinos with different masses at different speeds, so they are physical states. Their difference leads to the neutrino oscillations. Similarly, the spin states, in close correlative with the chirality states, construct the weak interaction Lagrangian density, so they can also be called the weak eigenstates, while the helicity states describe experimentally the polarization of fermions, so they are physical states. Their difference leads to the lifetime asymmetry.

The study of this difference can not only clarify many conceptual confusions and find a new polarization asymmetry phenomenon---the lifetime asymmetry, but also can predict the left-right polarization-dependent asymmetry of the weak interaction mass, which might be the reason why RH neutrinos can not be found experimentally \cite{19}.

Fermilab physicists hope to build a muon collider \cite{20}. The colliding of muons must happen before the muons decay. Therefore, it is necessary to consider the polarization asymmetry of muon lifetime in the design of a muon collider.

It might have consequences for cosmology and astrophysics. For example, the lifetime asymmetry could be served as one of mechanisms to explain the ``knee" and ``ankle" in cosmic ray spectrum \cite{21,22}.

\section*{acknowledgments}

We are grateful to Professor Guang-Jiong Ni and Professor V.~V.~Dvoeglazov for a stimulating discussion.

\end{document}